\documentclass[conference]{IEEEtran}
\IEEEoverridecommandlockouts

\usepackage[pdftex]{graphicx}
\usepackage{epstopdf}
\usepackage[utf8]{inputenc} 
\usepackage[T1]{fontenc}    
\usepackage{hyperref}       
\usepackage{url}            
\usepackage{booktabs}       
\usepackage{amsfonts}       
\usepackage{nicefrac}       
\usepackage{microtype}      

\usepackage{graphics, theorem, times, cite}

\usepackage{tikz}
 
\usepackage[english]{babel}
\usepackage{ifpdf}
\usepackage{bm}

\ifpdf
	\usepackage[pdftex]{graphicx}
	\graphicspath{{./figures/}}
\else
	\usepackage[dvips]{graphicx}
	\graphicspath{./figures/}
\fi
\usepackage{color}
\usepackage{pgf, tikz, pgfplots, epstopdf}
\pgfplotsset{compat=1.18}
\usetikzlibrary{shapes, arrows, automata}
\usepackage{subfig}
\usepackage{amsmath}
\usepackage{amsfonts, amssymb,  bbm}
\usepackage{mathrsfs}
\usepackage{upgreek}
\usepackage{algorithm,algpseudocode}
\usepackage{enumerate}
\usepackage{multirow}
\usepackage{rotating}
\usepackage{needspace}

\input{mySymbol.sty}

\usepackage{booktabs}       
\usepackage{nicefrac}       
\usepackage{microtype}      
\usepackage{xcolor}         

\newcommand{\QED}{\hfill\ensuremath{\blacksquare}}

\title{Wireless Link Scheduling with State-Augmented Graph Neural Networks  
}

\author{Romina Garcia Camargo \quad Zhiyang Wang \quad Navid NaderiAlizadeh \quad Alejandro Ribeiro \thanks{RGC, ZW, and AR are with the Department of Electrical and Systems Engineering, University of Pennsylvania, Philadelphia, PA (emails: \{rominag, zhiyangw, aribeiro\}@seas.upenn.edu). NN is with the Department of Biostatistics and Bioinformatics, Duke University, Durham, NC (email: navid.naderi@duke.edu).}}

\begin{document}

\maketitle

\begin{abstract}
We consider the problem of optimal link scheduling in large-scale wireless ad hoc networks. We specifically aim for the maximum long-term average performance, subject to a minimum transmission requirement for each link to ensure fairness. With a graph structure utilized to represent the conflicts of links, we formulate a constrained optimization problem to learn the scheduling policy, which is parameterized with a graph neural network (GNN). To address the challenge of long-term performance, we use the state-augmentation technique. In particular, by augmenting the Lagrangian dual variables as dynamic inputs to the scheduling policy, the GNN can be trained to gradually adapt the scheduling decisions to achieve the minimum transmission requirements. We verify the efficacy of our proposed policy through numerical simulations and compare its performance with several baselines in various network settings.

\end{abstract}


\section{Introduction}
\label{sec:intro}

{Large-scale wireless communication systems with increasingly complex architectures and diverse user demands have driven the need for efficient resource management strategies. One of the fundamental problems in wireless resource management is \emph{link scheduling}—deciding which links should transmit and when—to enhance overall network performance \cite{traditional3,traditional2,yi2015itlinq+,traditional4,spatialdl, grlinq}.} 
Wireless link scheduling plays a critical role in 
interference mitigation, fairness in resource allocation, and the overall quality of service (QoS) provided to end users.

Devising optimal link scheduling decisions is a fundamental challenge, especially in large-scale wireless networks. This challenge has been extensively studied, including from an information-theoretic perspective~\cite{jafar2013topological,naderializadeh2014interference}, resulting in a wide range of proposed solutions \cite{spatialdl, grlinq, GBlinksbeams, linkschedulingusinggnns, g-mlriemannian, GNNmultiuav}. 
{The work presented in \cite{spatialdl} proposes a deep neural network architecture that utilizes geographic location information of devices to optimize link scheduling decisions, effectively addressing interference challenges and improving network efficiency. Building upon a related scenario, the authors in \cite{grlinq} introduce an enhancement by employing K-nearest-neighbors (KNN) graph structures, enabling more accurate predictions of network performance based on local connectivity patterns. Further expanding the scope, the work in \cite{g-mlriemannian} presents a novel geometric machine learning approach leveraging Riemannian manifold structures. This method significantly reduces the required amount of training data by exploiting geometric relationships inherent in wireless network layouts, thereby enhancing scalability and adaptability. { In \cite{linkschedulingusinggnns}, the authors employ a graph neural network (GNN) framework to accelerate the computation of the Maximum Independent Set (MIS) in the network conflict graphs. These methods, particularly GNN-based models, naturally integrate conflict awareness into the scheduling process. Many existing studies adopt the instantaneous sum-of-rate as a performance metric. While effective in theoretical models, this instantaneous metric often does not accurately reflect real-world network dynamics. To better capture practical network behavior, both network-level and link-level transmissions, averaged over a long horizon, need to be considered as performance metrics. 
Such a joint consideration provides a more robust assessment of performance, ensuring minimum service guarantees for each link. }
}

In this work, we focus on learning an optimal link scheduling policy to maximize the long-term average performance of a large-scale ad-hoc wireless network, subject to meeting minimum long-term transmission constraints for each link. Our approach effectively identifies and mitigates conflicts among transmission links while explicitly ensuring fairness by enforcing minimum transmission thresholds per link, thereby facilitating sustained network efficiency over extended operational periods.

We propose a state-augmented wireless scheduling algorithm to optimize these time-averaged performance metrics. {Similar state augmentation strategies have been explored in \cite{navidsa,das2024learning,uslu2024learning} with a state-augmentation strategy to optimize performances over a long-term scale. Our contribution differs notably by incorporating state augmentation as a time-varying link scheduling strategy. Specifically, we formalize the link scheduling task as a constrained optimization problem, addressed through the Lagrangian dual framework. 
A traditional dual descent approach, however, does not necessarily guarantee the feasibility of the scheduling policies within a finite iteration horizon. As the dual variables capture how much the corresponding constraints are violated or satisfied, we incorporate them at each time instance as dynamic
inputs to the link scheduling policy. This enables us to train the policy to adapt its decisions to the minimum scheduling constraint satisfaction. 
Additionally, we use a GNN to parameterize the link scheduling policy, leveraging the underlying graph topology of the network to enable adaptive and effective scheduling decisions.
}

\section{Link Scheduling Problem Formulation}
\label{sec:problem}
\begin{figure}[h]
\centering
\includegraphics[width=0.8\linewidth]{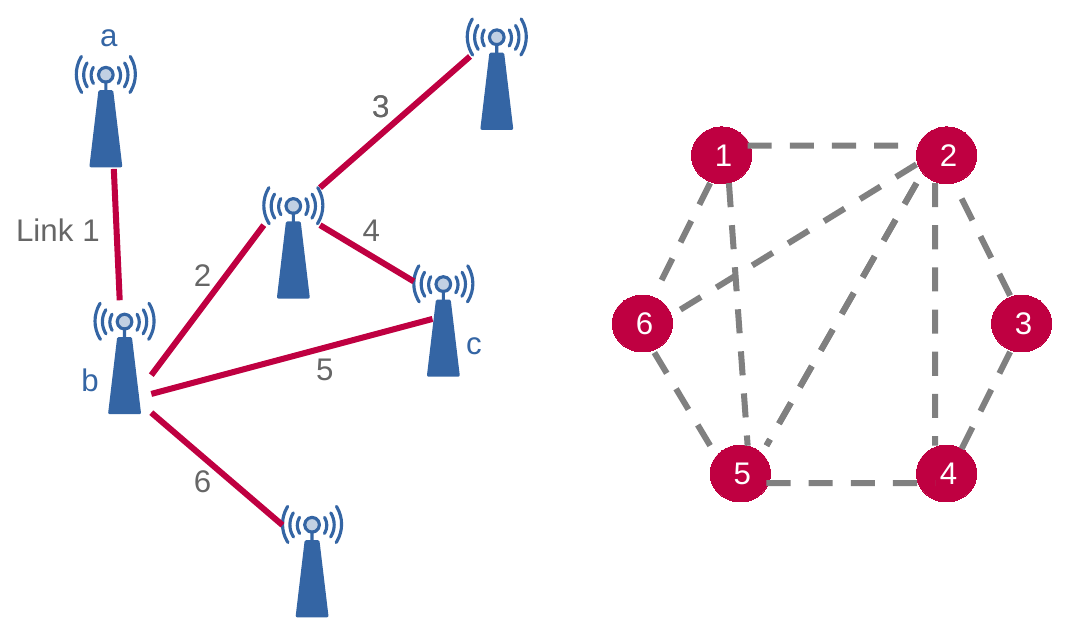}
\caption{An example wireless network. Left: Connectivity graph of the wireless network--The blue nodes represent the agents and the red edges mean the two agents can communicate with each other. Right: Conflict graph of the links in the wireless network-- The red nodes represent the links, while the edges indicate that the conflict exists because of the common agent. 
}
\label{fig:connect-interfere}
\end{figure}

We consider a wireless link scheduling problem over a series of time steps $t \in \{0, 1, 2, \dots, T-1\}$. The status for link $i\in\{1,2,\dots, K\}$ at time $t$ is denoted as $s_i(t)\in\{0,1\}$, where $s_i(t) = 1$ indicates that link $i$ is scheduled to transmit while $s_i(t) =0$ means no packet can be transmitted through link $i$.
The  relationships between links can be characterized by a conflict graph model $\bbG(\ccalV,
\ccalE)$, with the node set $\ccalV=\{1,2,\hdots, K\}$ representing the links. The edge set $\ccalE$ represents the relationships between each pair of links, i.e. $(i,j)\in \ccalE$ indicates a conflict exists between links $i$ and $j$ as they share a common agent. The unweighted adjacency matrix $\bbA\in\{0,1\}^{K\times K}$ denotes the conflict graph structure with the diagonal elements set as $0$.
With $\bbG$ as a model of the wireless network system, the scheduling status for each link $\bbs(t)=[s_1(t);s_2(t);\cdots; s_K(t)]\in\{0,1\}^K$ can be seen as the graph signal supported on the nodes $\ccalV$. 
Figure \ref{fig:connect-interfere} shows the communication graph and the link conflict graph in a small wireless network as an example.

Under this setting, the goal of the wireless link scheduling problem is to determine a scheduling policy under the conflict constraints imposed by the system architecture $\bbG(\ccalV,\ccalE)$, along with satisfying a minimum transmission requirement for each link, defined as $\bm\Delta\in \reals_+^{K}$. When the links are scheduled to transmit with policy $\{\bbs(t)\}_{t=0}^{T-1}$ over $T$ time slots, the network performance is measured by the average number of successful transmissions across $K$ links during $T$ time slots. The absence of simultaneous scheduling of conflicting links defines successful transmission. This is represented by the vector term $\mathbf{1}_K  - \bbA \bbs(t)$, with $1$ on the $i-$th entry indicating no conflicts when link $i$ is scheduled.  With these notations, we can formulate the optimal scheduling problem as the following constrained optimization problem.

\begin{align}
 P^* =&   \max\limits_{\substack{
        \{\bbs(t)\}_{t=0}^{T-1}, \\[2pt]
        \bbs(t)\in\{0,1\}^K
}}  \quad \frac{1}{T} \sum_{t=0}^{T-1} \bbs(t)^\mathsf{T}  [\mathbf{1}_K  - \bbA \bbs(t)]_+ \label{eq:objfunction} \\
s.t. & \quad \frac{1}{T} \sum_{t=0}^{T-1} \bbs(t)\odot[\mathbf{1}_K  - \bbA \bbs(t)]_+ \geq \bm\Delta, \label{eq:mintxconstrain}
\end{align}
and $\mathbf{1}_K$ is a vector of $K$ entries of value 1. Specifically, we denote $\odot$ as the element-wise product operation.

We note that heuristic and learning methods have been proposed to find the MIS over the conflict graph \cite{schneider2010optimal, zhao2023linksched}, while this cannot ensure each link is scheduled with a fair amount of transmissions. We implement a data-driven learning method to update the scheduling policy based on the previous observations. Consider the link scheduling problem as a constrained statistical learning problem, the scheduling policy can be substituted with a learnable function $\bm\Phi(\bbA, \bbH)$, which is a vector-valued function family with input as some parameter $\bbH\in \reals^s$ in a controllable dimension.

\begin{align}
 P' =&   \max\limits_{\bbH \in \reals^s}  \quad \frac{1}{T} \sum_{t=0}^{T-1} \bm\Phi(\bbA, \bbH)^\mathsf{T}  [\mathbf{1}_K  - \bbA \bm\Phi(\bbA, \bbH)]_+  \\
s.t. & \quad \frac{1}{T} \sum_{t=0}^{T-1} \bm\Phi(\bbA, \bbH)\odot[\mathbf{1}_K  - \bbA \bm\Phi(\bbA, \bbH)]_+ \geq \bm\Delta. 
\end{align}

This learnable function maps the wireless network status represented by the matrix $\bbA$ to the optimal scheduling policies. Though this parameterization provides a tractable and controllable solution domain, the output of this parameterized policy is the same for all times $t$, which we know is not an optimal solution to satisfy the time-averaging performance. We solve this issue with state-augmentation techniques that we introduce in the following section.

\section{State Augmented GNN Policies}
\label{sec:state_augment}
We first convert this constrained optimization problem presented in Eq. \eqref{eq:objfunction}-\eqref{eq:mintxconstrain} to the Lagrangian dual domain by importing a non-negative dual factor $\bm\lambda \in \reals_+^K$, which is used to penalize the violation of the minimum transmission constraint in Eq. \ref{eq:mintxconstrain}. The Lagrangian function can be explicitly written as
\begin{align}
& \overline{\ccalL}\left(\{\bbs(t)\}_{t=0}^{T-1}, \bm\lambda\right) = \frac{1}{T} \sum_{t=0}^{T-1} \left( \bbs(t)^\mathsf{T} [\mathbf{1}_K - \bbA \bbs(t)]_+ \right) \notag\\
& \quad \quad + \bm\lambda^\mathsf{T} \left( \frac{1}{T} \sum_{t=0}^{T-1} \bbs(t) \odot [\mathbf{1}_K - \bbA \bbs(t)]_+ - \Delta \right).
\end{align}
While we note that the variables can be decomposed with each time slot, we further decompose the Lagrangian function of the scheduling sequence as follows.
\begin{align}
\overline{\ccalL}\left(\{\bbs(t)\}_{t=0}^{T-1}, \bm\lambda\right)
&= \frac{1}{T} \sum_{t=0}^{T-1} \left( \bbs(t)^\mathsf{T} . [\mathbf{1}_K - \bbA . \bbs(t)]_+ \right. \notag\\
&  \left. + \bm\lambda^\mathsf{T} \left( \bbs(t)^\mathsf{T} \odot [\mathbf{1}_K - \bbA .\bbs(t)]_+ - \Delta\right) \right), \\
&  = \frac{1}{T} \sum_{t=0}^{T-1} \ccalL(\bbs(t), \bm\lambda).\label{eq:lagrangian}
\end{align}

To solve the problem in the dual domain, we need to maximize over the primal variables $\{\bbs(t)\}^{T-1}_{t=0}$, while then minimizing the dual variables $\bm\lambda$. In this case, the maximization of the primal variables is written as
\begin{align}
 \{\bbs^\star(t)\}_{t=0}^{T-1}& = \argmax_{\bbs(t)\in\{0,1\}^K}    \overline{\ccalL}(\{\bbs(t)\}_{t=0}^{T-1}, \bm\lambda), 
\end{align} 
which can be decomposed into each time step as 
\begin{align}
  \bbs^\star(t) & = \argmax_{\bbs(t)\in\{0,1\}^K}   {\ccalL}(\bbs(t), \bm\lambda).
\end{align} 

The dual variable optimization can be written as 

\begin{align}
  \bm\lambda^\star =  \argmin_{\bm\lambda\in \reals_+^K} \ccalL(\bbs^\star(t),\bm\lambda).
\end{align}
This can be approximated with gradient descent recursively as 
\begin{align}
\bm\lambda_{t+1} = \left[  \bm\lambda_t -\eta_{\bm\lambda} (\bbs(t)^\mathsf{T} \odot   [\mathbf{1}_K - \bbA \bbs(t)]_+  - \bm\Delta) \right]_+,
\end{align}
with $\eta_{\bm\lambda}$ representing the learning rate and $[\cdot]_+$ as the projection to the positive domain.

Theoretical results show that the proposed primal-dual scheduling policy can be both almost surely feasible and near-optimal when we operate over a long enough duration \cite{navidsa,ribeiro2010ergodic}. 
\begin{align}
    &\lim_{T\rightarrow \infty}\frac{1}{T} \sum_{t=0}^{T-1} \bbs(t)^\mathsf{T} [\mathbf{1}_K -\bbA \bbs(t)]_+ \geq P^* - \ccalO(\eta_{\bm\lambda})\\
    & \lim_{T\rightarrow \infty} \frac{1}{T}\sum_{t=0}^{T-1} \bbs(t)\odot[\mathbf{1}_K -\bbA \bbs(t)]_+ \geq \bm\Delta\qquad a.s.
\end{align}
These indicate that the scheduling policy satisfies the desired minimum transmission constraints and falls within a constant additive gap of the optimum $P^*$. 

However, these results hold only as the number of iterations, i.e. the operation time goes to infinity. This indicates that we may not obtain a feasible and near-optimal policy when we stop the training phase after a finite iterations. Furthermore, calculating the optimal policy for each dual variable $\bm\lambda_t$ can be costly, making the policy unrealizable. The lack of strong convexity in the linear function $\ccalL(\bbs(t),\bm\lambda)$ implies the optimal solution of $\overline{\ccalL}$ cannot be reached by optimizing the individual problem $\ccalL$.  

To address these challenges, we augment the scheduling policy with a corresponding set of dual variables $\bm\lambda$ as dynamic inputs which also indicate the constraint violation conditions \cite{naderializadeh2022state}. Furthermore, to represent the network policy with controllable and tractable solutions, we replace the scheduling policy with a GNN parameterization.

\subsection{GNN Parameterization}
We consider a Graph Neural Network (GNN) to parameterize our scheduling policy. GNNs consist of a cascade of layers, each containing a graph convolutional filter followed by a pointwise nonlinearity \cite{gama2019convolutional,scarselli2008graph}. Graph signals, which are values supported on nodes of the graph, can be seen as the information features of each link in the conflict graph. The graph shift operator \cite{ortega2018graph} is a graph matrix that diffuses graph signals over the edges of the graph. In our case, this is the adjacency matrix $\bbA$ in the link conflict graph. A graph convolutional filter is defined based on a consecutive graph shift operation that aggregates information of neighboring nodes. In the $l-$th layer with $\bbx_{l-1}$ as input graph signal, the graph convolutional filter can be written as
\begin{align}
    \bby_l = \sum_{k=0}^{K-1} h_{lk} \bbA^k \bbx_{l-1}, \qquad \bbx_{l-1}, \bby_l \in \reals^K,
\end{align}
where $\{h_{lk}\}_{k=0}^{K-1}$ is the filter parameters in the $l-$th layer. The output of graph filter is further passed through a point-wise nonlinearity operation $\sigma:\reals \rightarrow \reals$ as
\begin{align}
    \bbx_l = \sigma(\bby_l).
\end{align}
Given the evident similarity between a communication network and its graph representation, GNNs are commonly used as a mapping function from network scenarios to resource management strategies \cite{wang2022learning, eisen2020optimal, navidsa,shen2020graph}. 

There is growing evidence supporting the advantages of GNNs, which offer a wide range of desirable properties for real-world applications \cite{gnnprops, gama2019stability, ruiz2021transferability, wang2024geometric}. Notably, GNNs provide stability, a crucial feature for large-scale and dynamic systems often encountered in practical tasks. The architecture also ensures scalability, as the number of learnable parameters needed is independent of the system size.  Another key advantage of GNNs is their transferability: when trained on one graph, GNN can generalize effectively to different graph structures and signals generated from the same distribution \cite{ruiz2021transferability, wang2024geometric}. This helps to provide an alternative training method to tackle large-scale systems in practice with a better computational efficiency. 
Perhaps the most significant property of GNNs is their permutation equivariance, which they inherit from graph filters. This enables label-independent processing and allows GNNs to exploit the inherent symmetries of both graphs and graph signals. As a result, GNNs can learn multiple representations from a single input signal, enhancing the model's expressive power.

To leverage the benefits of GNNs, we parameterize our scheduling policy using this architecture, denoted by $\bm\Phi(\bbA, \bbH; \bbx)$ with the set of learnable parameters $\{h_{lk}\}_{lk}$ written as $\bbH\in\ccalH$ and $\bbx$ as the input graph signal.

\subsection{Proposed State Augmented GNN (SAGNN) Algorithm}
We augment the policy to have $\bm\Phi(\bbA, \bbH; \bm\lambda)$, such that the dual variable $\bm\lambda$ is the input graph signal varying over time. The use of state augmentation allows the GNN to account for the stochasticity in our problem. 

The augmented Lagrangian with a set of dual variables $\bm\lambda\in \reals^K_+$ and a parameterized GNN policy $\bm\Phi(\bbA,\bbH;\bm\lambda)$ is written as 
\begin{align}
 &\ccalL_{\bm\lambda}(\bbH)=  \bm\Phi(\bbA,
\bbH;\bm\lambda)^\mathsf{T}[\mathbf{1}_K - \bbA\bm\Phi(\bbA,
\bbH;\bm\lambda)]_+ \nonumber \\
&  + \bm\lambda^\mathsf{T} \left( \bm\Phi(\bbA,
\bbH;\bm\lambda)^\mathsf{T} \odot ([\mathbf{1}_K - \bbA \bm\Phi(\bbA,
\bbH;\bm\lambda)]_+ - \bm\Delta)\right)  \label{eq:L}
\end{align}

During the training phase we want to find the optimal model parameters for randomly drawn values of $\bm\lambda$. 
Considering $\bm\lambda \sim p_{\bm\lambda}$ with $p_{\bm\lambda}$ the probability distribution of the dual variable, the optimal state-augmented scheduling policy can be defined as
\begin{align}\label{opt:aug-primal}
\bbH^* & = \arg\max_{\bbH\in\ccalH} {\mathbb{E}_{\bm\lambda\sim p_{\bm\lambda}}} [\ccalL_{\bm\lambda}(\bbH)].
\end{align}
This can be solved iteratively with gradient ascent. Specifically, the training procedure of our proposed SAGNN is written in Algorithm \ref{alg:train}. We first randomly initialize the parameter as $\bbH_0$ and then update them over each iteration $n\in \{0,1,\cdots, N-1\}$ with a learning rate $\eta_{\bbH}$.
%

With the optimal state-augmented policy with $\bbH^*$ in \eqref{opt:aug-primal}, the Lagrangian-maximizing policy for each dual variable $\bm\lambda = \bm\lambda_t$. The state-augmented version of dual variable update is therefore 
\begin{align}\label{opt:aug-dual}
\nonumber &\bm\lambda_{t+1} = \Big[  \bm\lambda_t - \\
& \eta_{\bm\lambda} (\bm\Phi(\bbA,
\bbH;\bm\lambda_t)^\mathsf{T} \odot  \left(  [\mathbf{1}_K - \bbA\bm\Phi(\bbA,
\bbH;\bm\lambda_t)]_+  - \bm\Delta\right)) \Big]_+.
\end{align}

The execution phase consists of generating scheduling policies $\bm\Phi(\bbA,
\bbH^\star;\bm\lambda_t)$ when given the trained parameter $\bbH^\star$. The dual variable $\bm\lambda$ is updated every time step as \eqref{opt:aug-dual} shows, obtaining the optimal $\bm\lambda^\star$ as a result. 

The execution procedure of SAGNN is detailed in Algorithm \ref{alg:test}. The dual dynamics are employed to continuously monitor and adjust the satisfaction of the original constraints. Specifically, if the scheduling decisions at a given time step help satisfy the constraints, the dual variables decrease accordingly. Conversely, when the scheduling decisions result in constraint violations, the dual variables increase, guiding the scheduling policy back toward feasible solutions.


\begin{algorithm}
  \caption{Training Phase for the SAGNN Link Scheduling Algorithm}\label{alg:train}
  \begin{algorithmic}[1]
    \State{\textbf{Input:} Number of training iterations $N$, primal learning rate $\eta_{\bbH}$. 
    }
    \State{\textbf{Initialize:} $\bbH_0$ randomly}
      \For{$n=0,\cdots, N-1$}
      \State{Randomly sample $\bm\lambda\sim p_{\bm\lambda}$.}
      \State{Generate scheduling decisions $\bm\Phi(\bbA, \bbH_n;\bm\lambda)$.}
      \State{Calculate the augmented Lagrangian according to \eqref{eq:L}, i.e.
      \begin{align}
         & \ccalL_{\bm\lambda}(\bbH_n)=  \bm\Phi(\bbA,
\bbH_n;\bm\lambda)^\mathsf{T}[\mathbf{1}_K - \bbA\bm\Phi(\bbA,
\bbH_n;\bm\lambda)]_+ \nonumber \\
&  + \bm\lambda^\mathsf{T} \left( \bm\Phi(\bbA,
\bbH_n;\bm\lambda)^\mathsf{T} \odot ([\mathbf{1}_K - \bbA \bm\Phi(\bbA,
\bbH_n;\bm\lambda)]_+ - \bm\Delta)\right)
      \end{align} 
      }
 
        \State {Update the primal parameters
        $$ \bbH_{n+1} = \bbH_n + \eta_{\bbH} \nabla_{\bbH} \ccalL_{\bm\lambda}(\bbH_n).$$}
      \EndFor
      \State {\textbf{Return:} Optimal model parameters $\bbH^*.$}
  \end{algorithmic}
\end{algorithm}

\begin{algorithm}
  \caption{Execution Phase for the SAGNN Link Scheduling Algorithm}\label{alg:test}
  \begin{algorithmic}[1]
    \State{{\textbf{Input:} Optimal model parameters ${\bbH}^*$, number of time steps $T$, dual learning rate $\eta_{\bm\lambda}$}}
    \State{Initialize: $\bm\lambda_0 = \textbf{0}$}
      \For{$t=0,\cdots,T-1$}
      \State{Generate scheduling decisions $\bbs(t) = \bm\Phi(\bbA, \bbH^*;\bm\lambda_t)$.}
        \State{Update the dual parameters
        $$\bm\lambda_{t+1} = \left[  \bm\lambda_t -\eta_{\bm\lambda}  \nabla_{\bm\lambda}\ccalL_{\bm\lambda}(\bbH^*)
\right]_+.$$}

      \EndFor 
      \State \textbf{Return:} Optimal Scheduling decisions $\bbs(t)$, \\
      \quad $t=0,1,\cdots, T-1$.
  \end{algorithmic}
\end{algorithm}

\section{Numerical Experiments}
\label{sec:simu}
To evaluate the proposed algorithm, we run a series of experiments\footnote{The code used for the experiments can be found in \url{https://github.com/romm32/SAGNN}.}. The preliminary study focuses on constraint satisfaction under different transmission requirements, $\bm\Delta$, while also tracking both the count and the efficiency of scheduled transmissions. In this section, we denote $\bm\Delta = \Delta \mathbf{1}^K$, which means that we put the same minimum transmission requirement to each link.

Training and test data are created as follows. Communication topologies are instantiated as grid graphs—a structured subclass of random geometric graphs (RGGs). We place nodes on a regular lattice inside a bounded workspace and connect any two within a set communication radius. This grid layout offers a controlled yet scalable level of network complexity and structure. For each grid graph, we then compute its dual to obtain a conflict graph, giving a simplified but physically plausible model of conflict-constrained wireless communication networks.

We generate 60 communication graphs—10 for training and 50 for testing—using the following procedure. First, the number of nodes $N$ is drawn uniformly at random from the set $\{200,\dots,300\}$. The choice of the number of users is guided by the target number of links in the graph. A square workspace of side length $\sqrt{N}$ is then defined and partitioned into $\lceil\sqrt{N}\rceil \times \lceil\sqrt{N}\rceil$ square cells, each with side length $l_{\text{cell}} \;=\; \lceil \sqrt{N} \rceil .$
Each node is placed at the center of a cell, connecting two nodes whenever they fell within a radius of $1.2l_{cell}$ of each other. 
The communication graphs have an average degree of 4. This results in conflict graphs of medium size, with an average of $K\simeq 500$ links and an average node degree of 6.

Some heuristic baselines are considered for comparison with our proposed algorithm. We describe them as follows.

\begin{enumerate} [i.]
    \item \textit{p}-persistent scheduling: Determine a probability of transmission for each link, correlated to its degree of conflict. Sample Bernoulli trials according to these probabilities to schedule transmissions.
    \item Maximum independent set (MIS) scheduling: Consider the size of the MIS $M$, randomly choose $M$ nodes to transmit at each time step.
\end{enumerate}

For these baselines we consider a naive approach, where the successful transmissions are completely determined by our previous description, and a collision avoidance (CA) approach. The latter implies observing which conflicting links have been scheduled and randomly turning one of the two conflicting links off.

\subsection{Experimental setup}
The SAGNN model is trained for $N = 100$ epochs, with a batch size set to 1 to process graphs with different number of nodes independently. The minimum transmission requirement is not considered during training as it is a constant in the maximization of the Lagrangian in Eq. \ref{eq:L}. This approach allows the model to be trained independently of any specific constraint requirement, enhancing its generalization capability. Values for the dual variable $\bm\lambda$ are sampled from a uniform distribution $\mathcal{U}[0, 2]^K$, sampling 10 different vectors for each graph in each training epoch\footnote{For a more diverse set of vectors to be considered, we set $30\%$ of entries to 0 or $25\%$ of entries to 2 in some runs of the algorithm.
}. 

The Graph Neural Network (GNN) architecture consists of $L=3$ layers with 256 features each. TagConv \cite{tagconv} is used for the convolutional stages with filters of order $K=3$. Batch Normalization \cite{batchnorm} is applied to the output of the filters. The pointwise nonlinearity used is a leaky ReLU. The output of the model is passed through a sigmoid activation function to have values between 0 and 1. As the problem involves binary solutions, a threshold of 0.5 is used on evaluation to determine which links were scheduled for transmission. The primal learning rate, associated to the GNN's parameters, is set to $5\times10^{-5}$. Adam optimizer \cite{adam} is used to maximize the loss function defined in equation \ref{eq:L}, performing primal gradient ascent. 

Evaluation is carried out considering $T=200$ time steps. In the experiments, we use requirements varying in $\Delta=\{0.1, 0.125, 0.15\}$. The dual learning rate is set to 2. A resilient formulation of the problem \cite{resilience} is considered for the experiments to mitigate the effects that few links with infeasible constraints might have on the overall solution. The resilience factor is fixed at 0.05 for $\Delta=0.1$ and 0.1 for $\Delta=0.125, 0.15$.

\subsection{Basic performance of the SAGNN algorithm}
\begin{figure}
    \centering
\includegraphics[width=0.9\linewidth]{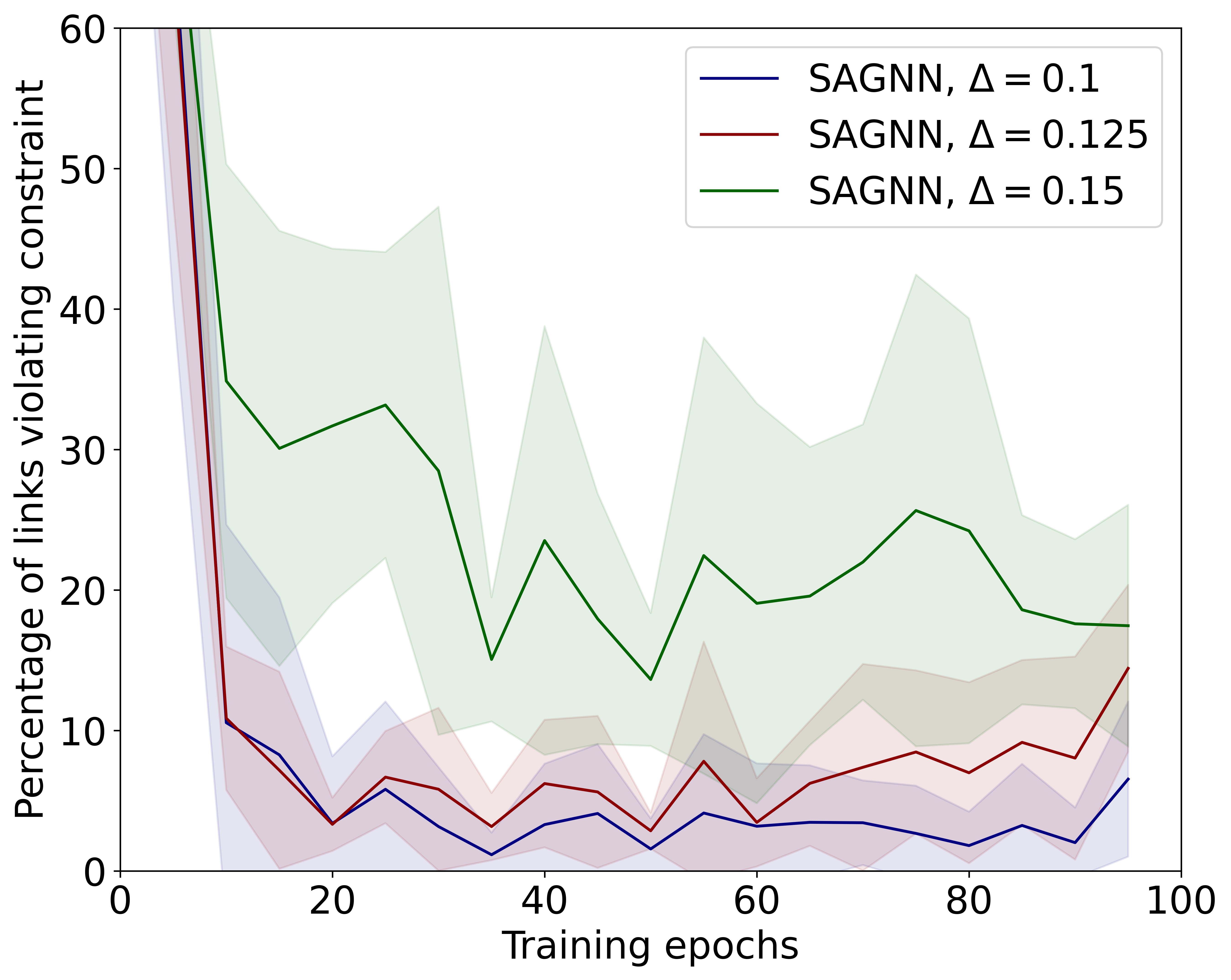}
    \caption{Evolution of constraint violation during training for different minimum transmission requirements. The mean across 3 experiments is presented, with the standard deviation shown in a lighter color. A running average of factor 5 was applied on the curves to enhance clarity.}
    \label{fig:constraints}
\end{figure}

We evaluate our algorithm after each training epoch to monitor its improvement throughout the training process. Figure \ref{fig:constraints} illustrates the average constraint violation during training, based on three independent experiments conducted on the test set. As training progresses, the level of constraint violation consistently decreases, indicating that the model is learning to better satisfy the minimum transmission constraint for each link. Additionally, we observe that higher values of the transmission requirement $\Delta$ make it more challenging to meet the constraints, reflecting increased problem complexity.

\begin{figure}
    \centering
    \includegraphics[width=0.9\linewidth]{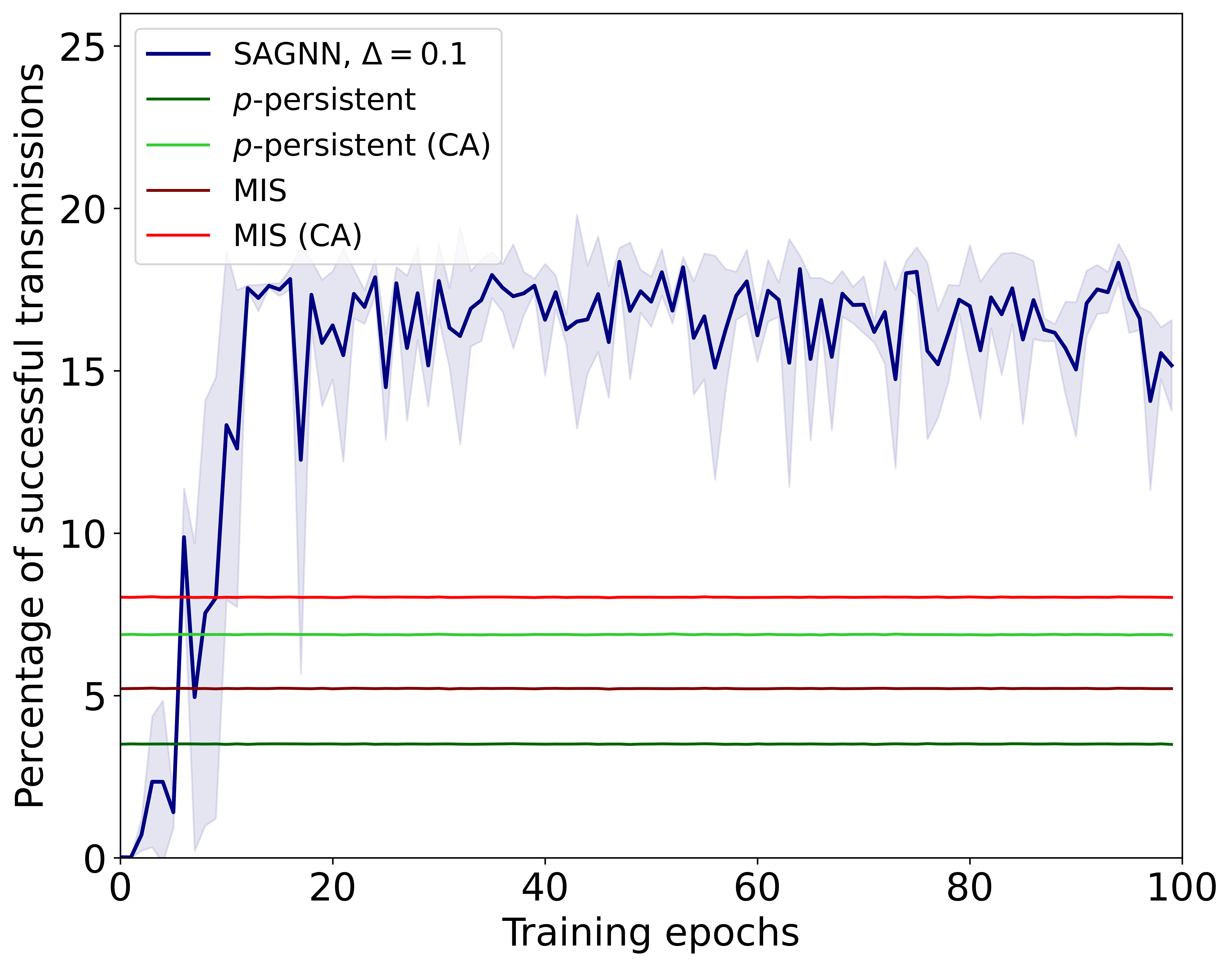}
    \caption{Evolution of the average successful transmissions in the network during training. The mean across 3 experiments is presented, with the standard deviation shown in a lighter color. Several baselines are added for comparison.}
    \label{fig:of}
\end{figure}

Figure \ref{fig:of} shows the value of the objective function, represented as the percentage of successfully active links in the network, considering a time average. This evaluation is carried out at each training epoch for 50 unseen graphs. Our proposed method, SAGNN, consistently achieves a higher number of successful transmissions on the test set compared to the baselines. It is important to note that the graphs used for evaluation have an average maximum independent set (MIS) size of approximately 25\% of the total links, which serves as an upper bound on the achievable value of the objective. However, this MIS upper bound does not consider the transmission constraints. Despite this, our model performs close to this theoretical limit, with the minimum transmission satisfied for each link, indicating strong performance.

\begin{figure}
    \centering
    \includegraphics[width=0.9\linewidth]{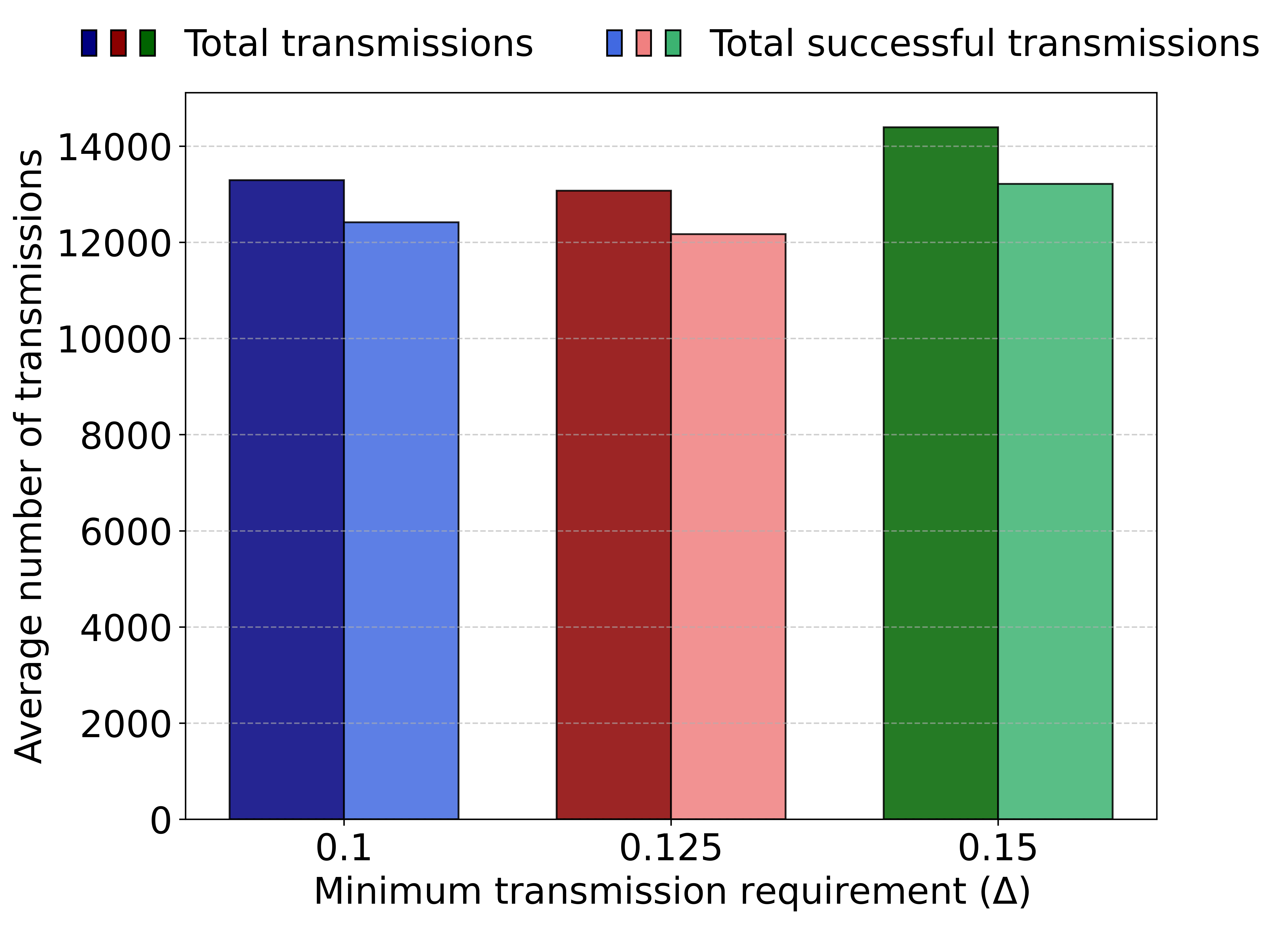}
    \caption{Number of total and successful transmissions for different requirements. The average across 100 graphs is presented for different values of $\Delta$.}
    \label{fig:plot8-succvstotal}
\end{figure}

In Fig. \ref{fig:plot8-succvstotal}, we compare the number of successful transmissions to the total number of transmission attempts. The algorithm is evaluated on 100 unseen graphs, and the results are averaged across these instances. We observe that the ratio of successful transmissions remains relatively stable as the minimum transmission requirement $\Delta$ increases. Notably, the algorithm learns to avoid scheduling conflicting links, which prevents wasted channel resources and contributes to the high success rate.

\subsection{Further analysis of constraint violation}
As shown in Fig. \ref{fig:constraints}, most links satisfy the constraints by the end of training. However, a small percentage still fall short of the transmission requirement. To better understand the extent of these violations, we perform a quantitative analysis using the following equation:

\begin{align}
    \text{violation level} = \left(\bm\Delta -\frac{1}{T} \sum_{t=0}^{T-1} \bbs(t)\odot[\mathbf{1}_K  - \bbA \bbs(t)]_+\right)\frac{1}{\bm\Delta}.
\end{align}

\begin{figure}
    \centering
    \includegraphics[width=0.9\linewidth]{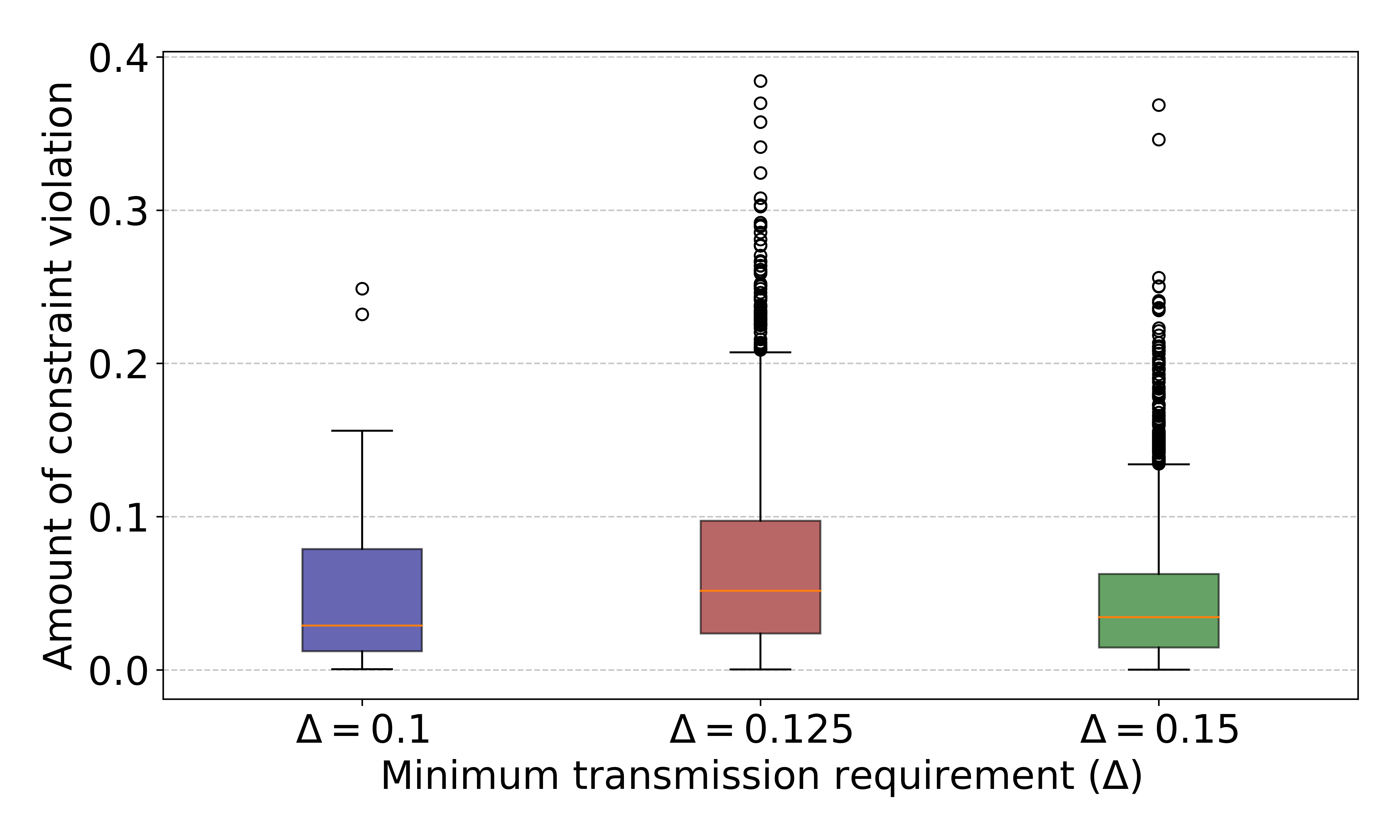}
    \caption{Box plot of the distribution of values for links violating the constraint. The values are scaled such that 1 is the largest possible value (which would indicate a complete violation of amount $\Delta$).}
    \label{fig:boxplot}
\end{figure}

Figure \ref{fig:boxplot} shows the distribution of constraint violation levels for links that fail to meet the transmission requirement. To facilitate comparison, we normalize the violations so that a value of 1 corresponds to a complete failure to transmit (i.e., a violation equal to $\Delta$). We observe that, for all values of $\Delta$ considered, the vast majority of violations remain below $10\%$, with only a few outliers. This provides strong evidence that our algorithm effectively learns to satisfy the constraints. The remaining violations may be attributed to instances where the problem is inherently infeasible for certain links.

\section{Conclusions}
\label{sec:conclusion}
We address the long-term link-scheduling problem in which each link must satisfy a minimum average transmission requirement. To learn an effective policy, we develop a State-Augmented Graph Neural Network (SAGNN) that incorporates the dual variables associated with the constraints as augmented, time-varying inputs. These dual variables indicate the current constraint violation, enabling the network to adapt its scheduling decisions and maintain feasibility over a long-term and time-averaging metric.
The proposed SAGNN is evaluated through comprehensive numerical experiments and is shown to outperform heuristic baselines. Looking ahead, we will tailor the minimum-transmission thresholds to reflect link-specific priorities, extend the framework to irregular graph topologies that better capture real-world deployments, and examine the stability and transferability of SAGNN across diverse network configurations. 


\urlstyle{same}




\end{document}